\documentclass[12pt]{article}
\usepackage{amsmath,amssymb}
\usepackage{graphicx,color}
\usepackage{cite}
\usepackage{bm}
\usepackage{dcolumn}
\def\be{\begin{equation}}
\def\ee{\end{equation}}
\def\bea{\begin{eqnarray}}
\def\eea{\end{eqnarray}}
\def\nn{\nonumber}

\newcommand{\Section}[1]{\section{#1}\setcounter{equation}{0}}
\renewcommand{\theequation}{\arabic{section}.\arabic{equation}}

\begin{document}

\pagestyle{plain}

\def\e{{\rm e}}
\def\cs{\frac{1}{(2\pi\alpha')^2}}
\def\CV{{\cal{V}}}
\def\haf{{\frac{1}{2}}}
\def\tr{{\rm Tr}}
\def\"{\prime\prime}
\def\p{\partial}
\def\tphi{\tilde{\phi}}
\def\ttheta{\tilde{\theta}}
\def\a{\alpha}
\def\b{\beta}
\def\la{\lambda}
\def\barla{\bar{\lambda}}
\def\ep{\epsilon}
\def\hj{\hat j}
\def\hn{\hat n}
\def\bz{\bar{z}}
\def\zk{{\bf{Z}}_k}
\def\h1{\hspace{1cm}}
\def\dd{\Delta_{[N+2k] \times [2k]}}
\def\ddbar{\bar{\Delta}_{[2k] \times [N+2k]}}
\def\u{U_{[N+2k] \times [N]}}
\def\ubar{\bar{U}_{[N] \times [N+2k]}}
\def\goes{\rightarrow}
\def\goal{\alpha'\rightarrow 0}
\def\ads2{AdS_2 \times S^2}
\def\ola{\overline {\lambda}}
\def\oep{\overline {\epsilon}}

\vspace{3cm}

\title{\bf $AdS/CFT$ correspondence via R-current correlation functions revisited}
\author{Shahin Mamedov$^{1,3}$\thanks{Email: sh$_-$mamedov@yahoo.com \hspace{2mm} \&
\hspace{2mm}  shahin@theory.ipm.ac.ir } \hspace{3mm} and
\hspace{3mm} Shahrokh Parvizi$^{2,3}$\thanks{Email:
\hspace{2mm}  parvizi@theory.ipm.ac.ir }\\
{\small {\em 1. Institute for Physics Problems, Baku State University, }}\\
{\small {\em Z. Khalilov str.23, AZ-1148, Baku, Azerbaijan}}\\
{\small {\em 2. Department of Physics, Sharif University of Technology}} \\
{\small {\em P.O. Box 11365-9161, Tehran, IRAN}}\\
{\small {\em 3. Institute for Studies in Theoretical Physics and
Mathematics (IPM),}} \\
{\small {\em P.O.Box 19395-5531, Tehran, Iran}}
 } \maketitle

\begin{abstract}
\noindent Motivated by realizing open/closed string duality in the
work by Gopakumar [Phys. Rev. D70:025009,2004], we study two and
three-point correlation functions of R-current vector fields in
${\cal N}=4$ super Yang-Mills theory. These correlation functions
in free field limit can be derived from the worldline formalism
and are written as heat kernel integrals in the position space. We
show that reparametrising these integrals converts them to the
expected $AdS$ supergravity results which are known in terms of
bulk to boundary propagators. We expect that this
reparametrization corresponds to transforming open string moduli
parametrization to the closed string ones.
\end{abstract}

\vspace{1cm} PACS: 11.25.Tq

\vspace{3cm}
%\begin{flushright}
IPM/P-2005/028

SUT-P-06-1b

hep-th/0505162
%\end{flushright}

%%%%%%%%%%%%%%%%%%%%%%%%%%%%%%%%%%%%%%%%%%%%%%%%%%%%%%%%%%%%%%%%%%%%
\newpage

\hspace{15mm} \newline {\Large {\bf Introduction}} \vspace{4mm}
\newline \noindent The early idea of $AdS/CFT$ correspondence is
based on the duality between the supergravity theory in the bulk
of $AdS$ spacetime on one side and ${\cal N}=4$ super Yang-Mills
theory living on the boundary on the other. An important
realization of this correspondence is the derivation of the
correlation functions of the boundary theory from the partition
function of the bulk theory \cite{1,2}. In this connection the
correlation functions in $AdS/CFT$ have been studied by different
authors \cite{3}-\cite{Moghimi-Araghi:2004ds}. However, it is
believed that this duality has a deeper origin in the underlying
string theories. In other words, the  $AdS/CFT$ duality is a
consequence of the closed/open string duality. The SYM theory on
the boundary and supergravity in the bulk are effective theories
of open and closed strings, respectively. Despite of efforts in
understanding this duality in the level of string theories in
diverse aspects (e.g. \cite{Berenstein:2002jq}), an interesting
idea would be revisiting correlation functions and thinking how
they could be realized as an closed/open string duality. In this
regard, a worth asking question is whether it is possible to
somehow glue up the open string amplitudes corresponding to
correlators on the boundary to find the closed string amplitudes
in the bulk. The closed/open string dualities were observed in
some specific examples in the context of the topological string
theory in \cite{Gopakumar:1998ki}-\cite{Gaiotto:2003yb}. However
an affirmative clear answer to this question, in the original
$AdS/CFT$ conjecture, comes in a series of elegant works by
Gopakumar in \cite{16,16.2,16.3} where the basic tool is the
worldline formalism in the limit of weak coupling or free fields
(see also \cite{Akhmedov:2004yb}).

The worldline formalism was developed for calculation of field
theory correlators at one loop approximation \cite{14,15}. This
formalism is based on the idea of converting the field theory path
integrals in one-loop effective action into integrals over the
parameters defined at one loop which turns out to  be a
realization of the open string moduli. The worldline formalism has
found successful applications in field theory problems related to
the one-loop effective action \cite{15,27}. In \cite{16}, the
worldline formalism was found useful as well for rewriting
correlators in terms of open string worldsheet moduli parameters.
This is motivated by the fact that this parametrization comes
directly from open string theory in $\a'\goes 0$ limit. Then an
analogy with the electrical networks suggests a reparametrization
known as \emph{delta} to \emph{star} which corresponds to a
transformation which converts an open string one loop world sheet
to a closed string sphere with holes replaced by closed string
vertices. After this reparametrization, the amplitude can be
understood as an amplitude in $AdS$ spacetime which is constructed
from bulk to boundary (as well as bulk to bulk) propagators in the
$AdS$ spacetime. In a free scalar theory the two and three point
correlators in a worldline formulation were written in the form of
heat kernel integrals. After some reparametrization, such
integrals obey scalar field equation in $AdS$ spacetime, i.e. they
are the bulk to boundary propagators in this spacetime. The one
advantage of this  approach is to observe explicitly the footprint
of  $AdS$ bulk to boundary propagators in field theory
correlators. It realizes the $AdS/CFT$ correspondence at the level
of free string amplitudes.

%%%%%%%%%%%%%%%%%%%%%%%%%%%%%%%%%%%%%%%%%%%%%%%%%%%%%%%%%%%%%%%%
\begin{figure}\label{fig1}
 \begin{center}
  \includegraphics[scale=.6]{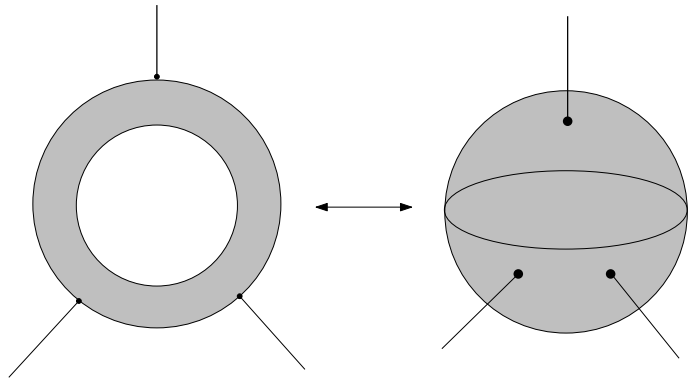}
 \end{center}
\caption{An open string one-loop amplitude (annulus) is dual to a
closed string tree diagram (sphere). }
\end{figure}

%%%%%%%%%%%%%%%%%%%%%%%%%%%%%%%%%%%%%%%%%%%%%%%%%%%%%%%%%%%%

One may think that the magic of deriving the $AdS$ spacetime
structure from a free field  theory comes from the simple
structure of scalar fields, while in both ${\cal N}=4$ SYM and
Supergravity theories, we have more complicated objects (even in
free limit) and investigation of the duality via correlators
remains non-obvious. An interesting early work on the
correspondence of non-obvious correlators is \cite{21} (see also
\cite{Osborn:1993cr}), where two and three point correlators of
R-symmetry currents in ${\cal N} =4$ super Yang-Mills theory were
obtained and confirmed the $AdS/CFT$ correspondence. In this
paper, we try to implement the ideas of \cite{16} to R-currents,
i.e. in the worldline formalism framework we use the procedure of
introducing bulk to boundary propagators via parameter integrals
of heat kernel in the two and three point $R$-current correlation
functions in super Yang-Mills theory. The  importance of
$R$-currents, besides their rich structure as vector fields, comes
from the fact that they, as well as scalar fields, are protected
objects in ${\cal N}=4$ SYM theory by supersymmetry. This enables
us to use them in the free field limit and trust our results in
wider limits. However, we have to restrict ourselves to at most
three-point function, since higher n-point functions require bulk
to bulk propagators in the $AdS$ space which in turn requires the
sum over all string states. In contrast, in three-point function
the only relevant propagator is the bulk to boundary propagator
and it is enough to consider the lowest string state for this
propagator. The important result is that when one considers the
one loop correlation in the SYM theory, it is possible to
reparametrise it in the same fashion of \emph{delta} to
\emph{star} in the electrical networks and find out an amplitude
which corresponds to a tree diagram. This amplitude is of course
the closed string sphere diagram and explicitly can be shown to
correspond to an amplitude of the bulk theory.

This paper is organized as follows.  In section 1 we introduce the
R-currents in $SYM$ theory and use the basic results of worldline
formalism to derive the two point functions. Then in section 2 we
investigate the three point function. In section 3 we introduce
the vector field theory in the $AdS$ theory and review the
derivation of the R-current correlation functions from the bulk
theory. Section 4 is devoted to bringing the results of worldline
formalism and $AdS$ supergravity in agreement with each other and
we conclude in section 5. A very brief review on the worldline
formalism is given the appendix.

%%%%%%%%%%%%%%%%%%%%%%%%%%%%%%%%%%%%%%%%%%%%%%%%%%%%%%%%%%%%%%%%%%%%%%%%%%%%%%
\Section{R-currents in the SYM theory:two-point correlation
functions}

Let us start with the ${\cal N}=4$ super Yang-Mills theory which
is given by the following action: \bea S&=&\tr \int d^4x
\frac{1}{4} F_{\mu\nu}F^{\mu\nu} -\frac{i}{2}\overline{\psi}_a
\gamma^\mu D_\mu\psi^a -\haf
D_\mu\phi_{ab}D^\mu\phi^{ab}  \nn\\
&& -\frac{i}{2} \overline{\psi}_a [\phi^{ab},\psi_a] + \frac{1}{4}
[\phi_{ab},\phi_{cd}][\phi^{ab},\phi^{cd}] \; , \eea where $\mu,
\nu=1,...,4$ are 4-dimensional space-time indices and
$a,b=1,...,6$ are the internal directions. Under the R-symmetry
fermions transform chirally in the fundamental representation {\bf
4} of $SU(4)$ and scalars transform as antisymmetric {\bf 6},
$$  \delta \psi^a = \ep^A(T_A)^a_b \frac{(1+\gamma_5)}{2}\psi^b ,
\;\;\;\; \delta \phi_{ab}=\ep^A(T_A)_{ab}^{cd} \phi_{cd}  $$

Then the corresponding global $SU(4)$ R-currents can be derived
as, \bea J^\mu_a = \haf \phi(x) T_a^{\phi} ( \partial^\mu +
A^\mu(x))\phi(x)-\frac{i}{2} \overline{\psi}(x)T_a^{\psi}
\gamma^\mu \frac{(1+\gamma_5)}{2}\psi(x).
 \eea

Our aim is to find two and three point correlation functions for
these R-currents which will be later interpreted as open string
amplitude then in the following sections we will show that these
can be transformed  to closed string amplitudes. This
transformation would be a realization of the open/closed string
duality. In this regard, it is helpful to use the worldline
formalism which in turn is a stringy inspired method in quantum
field theory \cite{15}. Indeed this formalism is the first
quantization (in contrast to second quantization in the field
theory) and so is comparable to the perturbative string theory.

The one-loop $N$-point R-current amplitude has contributions from
both scalar and spinor loops as demonstrated in the diagram of
Fig. 2.

%%%%%%%%%%%%%%%%%%%%%%%%%%%%%%%%%%%%%%%%%%%%%%%%%%%%%%%%%%%%%%%%
\begin{figure} \label{fig2}
 \begin{center}
  \includegraphics[scale=.6]{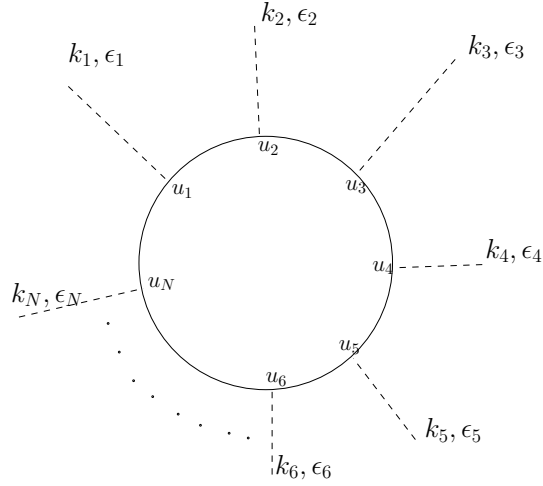}
 \end{center}
\caption{The Feynmann diagram for $N$ point function of vector
fields (dashed lines) evaluated by a scalar or spinor loop.
$k_i$'s and $\epsilon_i$'s are the momenta and polarizations of
vector fields, respectively. $u_i$'s are the insertion points on
the worldline loop. }
\end{figure}

%%%%%%%%%%%%%%%%%%%%%%%%%%%%%%%%%%%%%%%%%%%%%%%%%%%%%%%%%%%%

The scalar loop contribution to the $N$-point amplitude in
$d=4-\epsilon$ dimension can be derived in the worldline formalism
as \cite{14}:\footnote{see the appendix for a brief derivation.}
$$
\Gamma _N\left( k_1,...,k_N\right) =\frac{\left( ig\right)
^N}{\left( 4\pi
\right) ^2}Tr\left( T^{a_N}...T^{a_1}\right) \int\limits_0^\infty \frac{d\tau}{%
\tau^{3-N-\epsilon/2}}\int\limits_0^1du_{N-1}\int\limits_0^{u_{N-1}}du_{N-2}\dots\int\limits_0^{u_2}du_1
$$
\begin{equation}
\label{28}\times \exp \left[ \sum\limits_{i<j=1}^N\left( k_i\cdot
k_jG_B^{ji}-i\left( k_i\cdot \epsilon _j-k_j\cdot \epsilon
_i\right)
\stackrel{.}{G}_B^{ji}+\epsilon _i\cdot \epsilon _j\stackrel{..}{G}%
_B^{ji}\right) \right] \mid _{linear\ in\ each\ \epsilon }.
\end{equation}
Here $\tau$ is the Schwinger proper-time and $u_i$ are parameters
ordered in worldline loop $u_N\geq u_{N-1}\geq...\geq u_1$,
$\epsilon _i,$ and $k_i,$ are polarization vectors and momenta of
incoming and outgoing vector fields (gluons). The order of color
$T^a$ matrices under the trace should be the same as the order of
$u_i$ parameters, which is determined by positions of vector
currents on the loop.

Spinor loop contribution, for the above ordering, is given by:
$$
\Gamma _N\left( k_1,...,k_N\right) =-2\frac{\left( ig\right)
^N}{\left( 4\pi
\right) ^2}Tr\left( T^{a_N}...T^{a_1}\right) \int\limits_0^\infty \frac{d\tau}{%
\tau^{3-N-\epsilon /2}}\int\limits_0^1du_{N-1}\int%
\limits_0^{u_{N-1}}du_{N-2}...\int\limits_0^{u_2}du_1
$$
$$
\times \exp \left[ \sum\limits_{i<j=1}^Nk_i\cdot
k_jG_B^{ji}\right] \left\{ \prod\limits_{i=1}^N\int d\theta
_id\overline{\theta }_i\exp \left[ \sum\limits_{i<j=1}^N\left(
-i\left( \overline{\theta }_j\theta _jk_i\cdot \epsilon
_j-\overline{\theta }_i\theta _ik_j\cdot \epsilon _i\right)
\stackrel{.}{G}_B^{ji}\right. \right. \right.
$$
\begin{equation}
\label{30}\left. \left. \left.+ \overline{\theta }_i\theta _i\overline{%
\theta }_j\theta _j\epsilon _i\cdot \epsilon _j\stackrel{..}{G}%
_B^{ji}\right) +\left( -\overline{\theta }_i\overline{\theta
}_jk_i\cdot
k_j+i\overline{\theta }_i\theta _jk_i\cdot \epsilon _j+i\theta _i\overline{%
\theta }_j\epsilon _i\cdot k_j+\theta _i\theta _j\epsilon _i\cdot
\epsilon _j\right) G_F^{ji}\right] \right\}
\end{equation}
where $\theta ,\overline{\theta }$ are Grassmann variables and
bosonic and fermionic worldline Green's functions are defined in
the loop as:
$$
G_B^{ji}\equiv\left( \left| \tau_j-\tau_i\right| -\frac1\tau
\left( \tau_j-\tau_i\right) ^2\right)  =\tau\left( \left|
u_j-u_i\right| -\left( u_j-u_i\right) ^2\right) ,
$$
$$
G_F^{ji}=sign\left( \tau_j-\tau_i\right),
$$
with $\tau_i=u_i \tau$. Notice, that formulas (\ref{28}) and
(\ref{30}) do not contain the self interaction of vector field,
which leads to one-particle reducible diagrams. So, here we are
going to consider only contribution of one-particle irreducible
diagrams. The first and second derivatives of worldline Green's
function $G_B^{ji}$ are: \bea \frac \partial {\partial
\tau^j}G_B^{ji}= \stackrel{.}{G}_B^{ji}= sign \left(
\a^j-\a^i\right) \left( 1-2\left| \epsilon ^{ijk}\right| \alpha
_k\right) =\left( -1\right) ^{F_{ij}}\left( 1-2\left| \epsilon
^{ijk}\right| \alpha _k\right) , \eea \bea \label{32}\frac
{\partial^2} {\left(\partial \tau^j\right)^2}G_B^{ji}=
\stackrel{..}{G}_B^{ji}=\frac 2\tau \left( \left| \epsilon
^{ijk}\right| \delta \left( \alpha _k\right) -1\right) , \eea Here
$\a_i=|u_i-u_{i+1}|$, $\epsilon^{ijk}$ is unit antisymmetric
tensor and $\left(-1\right)^{F_{ij}}$ replaces the signature
function above:  $F_{ij}=\left\{
\begin{array}{c}
1\ for\ j<i \\
0\ for\ j>i
\end{array}
\right.$.

Beside the scalar and spinor loops, in super Yang-Mills theory
there is a vector field loop.  However since we are working in the
free field limit, it doesn't contribute.

Now let us concentrate on two-point function of vector fields
which is known in literature as polarization operator (photon or
gluon). Worldline expression of scalar loop contribution to this
function can be obtained from (\ref{28}), with $N=2$. After little
simplification it gets the following form, which coincides with
the ordinary expression in Schwinger's proper time parameter
\cite{14}:

\bea\label{34} \Pi^{ab}_{{\rm scalar}} \left( k_1,k_2\right)
&=-&Tr\left(T^a T^b\right) \frac{\left( g\mu ^{\epsilon/2}\right)
^2}{\left( 4\pi \right) ^{2-\epsilon/2}}\int\limits_0^\infty
\frac{d\tau }{\tau ^{d/2-1}} \int\limits_0^1d\alpha \ e^{\tau
k_1k_2\alpha \left( 1-\alpha \right)} \int\limits_{-\infty
}^\infty \frac{d^dz}{\left( 2\pi \right) ^d}e^{i\left(
k_1+k_2\right) z}   \nn\\
&&\times \left[ \frac 2\tau \left( \delta \left( \alpha \right)
-1\right) \epsilon _1\cdot \epsilon _2+\left( 1-2\alpha \right)
^2\left( \epsilon _1\cdot k_2\right) \left( \epsilon _2\cdot
k_1\right) \right] \; . \eea The last integral means the
energy-momentum conservation. Remind that the polarization
operator (\ref{34}) contains contribution of tadpole diagram of
vector-scalar interaction in $\delta \left( \alpha
\right)$-function term. It differs from two-point correlation
function of free scalar field only by additional square bracket
factor in it \cite{14,13}. Now the next step is going to position
space by means of the inverse Fourier transformation: \bea
\Pi^{ab}_{{\rm scalar}} \left( x_1,x_2\right)
=-Tr\left(T^aT^b\right) C\int\limits_0^\infty \frac{d\tau }{\tau
^{d/2-1}} \int\limits_{-\infty }^\infty d^dk_1
\int\limits_{-\infty }^\infty d^dk_2 \int\limits_{-\infty }^\infty
\frac{d^dz}{\left( 2\pi \right) ^d}e^{i\left(
k_1+k_2\right)z}\textsl{}
\nn \\
\times\int\limits_0^1d\alpha \ \int\limits_0^1d\beta \ e^{-\tau
k_1^2\beta \alpha \left( 1-\alpha \right) -\tau k_2^2\left(
1-\beta
\right) \alpha \left( 1-\alpha \right) } e^{-ik_1x_1-ik_2x_2} \nn\\
\label{36} \times \left[ \frac 2\tau \left( \delta \left( \alpha
\right) -1\right) \epsilon _1\cdot \epsilon _2+\left( 1-2\alpha
\right) ^2\left( \epsilon _1\cdot \left( i\frac \partial {\partial
x_2}\right) \right) \left( \epsilon _2\cdot \left( i\frac \partial
{\partial x_1}\right) \right) \right] \eea

Here we have inserted an integral over the parameter $\beta$,
where $0\leq \beta\leq 1$. Taking Gaussian integrals over the
momenta and then position derivatives, we obtain from (\ref{36})
the following expression of polarization operator\footnote{We have
included constants into the new one.}: \bea \label{38}
\Pi^{ab}_{{\rm scalar}} \left( x_1,x_2\right)
&=&-CTr\left(T^aT^b\right)\int\limits_0^\infty \frac{d\tau }{\tau
^{d/2-1}}\int\limits_0^1d\alpha \ \int\limits_0^1d\beta \
\int\limits_{-\infty }^\infty d^dz\  \left( \frac \pi
{\tau \beta \alpha \left( 1-\alpha \right) }\right) ^{d/2} \nn\\
&& \times e^{-\frac{ \left( z-x_1\right) ^2}{4\tau \beta \alpha
\left( 1-\alpha \right) }}\left( \frac \pi {\tau \left( 1-\beta
\right) \alpha \left( 1-\alpha \right) }\right)
^{d/2}e^{-\frac{\left( z-x_2\right) ^2}{4\tau \left( 1-\beta
\right) \alpha \left( 1-\alpha \right) }} \left[ \frac 2\tau
\left( \delta \left( \alpha \right) -1\right) \epsilon _1\cdot
\epsilon
_2 \right.\nn\\
&&\left. -\left( 1-2\alpha \right) ^2\frac 1{2\tau \left( 1-\beta
\right) \alpha \left( 1-\alpha \right) }\epsilon _1\cdot \left(
z-x_2\right) \frac 1{2\tau \beta \alpha \left( 1-\alpha \right)
}\epsilon _2\cdot \left( z-x_1\right)
\right] .\nn\\
\eea Writing the exponents in heat kernel terms by means of:
\begin{equation}
\label{40}\left( \frac 1{4\pi t}\right) ^{d/2}\ e^{-\frac{\left(
x-z\right) ^2}{4t}}=\left\langle x\left| e^{t\Box }\right|
z\right\rangle ,
\end{equation}
we rewrite (\ref{38}) as follows:
\begin {equation}
\label{42} \Pi^{ab}_{{\rm scalar}} \left( x_1,x_2\right)
=-C^{\prime }Tr\left(T^aT^b\right)\int\limits_0^\infty \frac{d\tau
}{\tau ^{d/2-1}}\int\limits_0^1d\alpha \ \int\limits_0^1d\beta \
\int\limits_{-\infty }^\infty d^dz \left\langle x_1\left| e^{\tau
\beta \alpha \left( 1-\alpha \right) \Box }\right|
z\right\rangle\times\nn\\
$$
$$
\left\langle x_2\left| e^{\tau \left( 1-\beta \right) \alpha
\left( 1-\alpha \right) \Box }\right| z\right\rangle\left[ \frac
2\tau \left( \delta \left( \alpha \right) -1\right) \epsilon
_1\cdot \epsilon _2-\left( 1-2\alpha \right) ^2\frac{\epsilon
_1\cdot \left( z-x_2\right) }{2\tau \alpha\left( 1-\beta
\right)\left(1-\alpha \right) } \frac{\epsilon _2\cdot \left(
z-x_1\right) }{2\tau \beta \alpha \left(
1-\alpha \right) }\right].\nn\\
\end{equation}

We can insert the integral representation of $\Gamma \left(
s\right) $ function,
\begin{equation}\label{46}
1=\frac 1{\Gamma \left( d/2\right) }\int\limits_0^\infty d\rho
\rho ^{d/2-1}e^{-\rho }=\frac 1{\Gamma \left( d/2+1\right)
}\int\limits_0^\infty d\rho \rho ^{d/2}e^{-\rho },
\end{equation}
into (\ref{42}) and pass to new variables \cite{16}:
$$\rho _1=\rho
\left( 1-\beta \right) ,\ \rho _2=\beta \rho \;\;\; {\rm and}
\;\;\; t=4\tau \rho \beta \left( 1-\beta \right) \alpha \left(
1-\alpha \right). $$ with $\int\limits_0^\infty \rho d\rho
\int\limits_0^1d\beta =\int\limits_0^\infty d\rho
_1\int\limits_0^\infty d\rho _2$, then  the polarization operator
(\ref{42}) is rewritten in a more symmetric form:
\begin{equation}
\label{48} \Pi^{ab}_{{\rm scalar}} \left(
x_1,x_2\right)=\int\limits_0^\infty \frac{dt}{t^{\frac{d}{2}+1}}\
\int\limits_{-\infty }^\infty d^dz\int\limits_0^\infty d\rho
_1\rho _1^{\frac{d}{2}-1}e^{-\rho _1}\left\langle x_1\left|
e^{\frac t{4\rho _1}\Box }\right| z\right\rangle
\int\limits_0^\infty d\rho _2\rho _2^{\frac{d}{2}-1}e^{-\rho
_2}\left\langle
x_2\left| e^{\frac t{4\rho _2}\Box }\right| z\right\rangle \nn \\
$$
$$
\times\left[ -\frac{C^{ab}_1}{\Gamma \left(d/2+1\right) }t\epsilon
_1\cdot \epsilon _2+\frac{C^{ab}_2}{\Gamma \left(d/2\right)
}\epsilon _1\cdot \left( z-x_2\right) \epsilon _2\cdot \left(
z-x_1\right) \right] .
\end{equation}

Here we have taken into account $\rho _1+\rho _2=\rho $ in the
exponent and have
included integrals over the $\alpha $ into constants $C_{1,2}^{ab}$%
\footnote{We suppose $d\geq 2$ for convergence of these integrals.
Finally, we put $d=4$.}:
$$
\begin{array}{c}
C^{ab}_1=Tr\left(T^aT^b\right)4^{d/2-1}C^{\prime
}\int\limits_0^1d\alpha \ \left[ \alpha \left( 1-\alpha \right)
\right] ^{d/2-1}\left( \delta \left( \alpha \right)
-1\right) , \\
C^{ab}_2=Tr\left(T^aT^b\right)4^{d/2-2}C^{\prime
}\int\limits_0^1d\alpha \ \left[
\alpha \left(1-\alpha \right) \right] ^{d/2-2}\left( 1-2\alpha \right) ^2,\\
C^{\prime }=g^2\mu ^{4-d}2^{d/2-6}\left( 2\pi \right) ^{d/2-4}
\end{array}
$$
Note that , the $\Gamma \left( s\right) $ function arises here
naturally, as a well-known correction (using generalized $\zeta -$
function in \cite{17} and other approaches) to the one-loop
effective action. Following \cite{16} we separate the integrals
over the $\rho_i$, which contain heat kernel, and denote them by
the $K_1\left( x_i,z,t\right)$ function:
\begin{equation}\label{50}
K_1\left( x_i,z,t\right) =\int\limits_0^\infty d\rho _i\rho
_i^{d/2-1}e^{-\rho _i}\left\langle x_i\left| e^{\frac t{4\rho
_i}\Box }\right| z\right\rangle.
\end{equation}
Given the $AdS$ metric as $ds^2=(dz_0^2+d\vec{z}^2)/z_0^2$, it can
be shown that the
function $K_1\left( x_i;z,t\right) $ obeys the $d+1$%
-dimensional massless Klein-Gordon equation in the $AdS$
background:
\begin{equation}
\label{52}\left[ -z_0^2\partial _{z_0}^2+\left( d-1\right)
z_0\partial _{z_0}-z_0^2\Box \right] K_1\left( x;z,t\right) =0,
\end{equation}
where $t=z_0^2$ and $\Box $ is the $d$-dimensional Laplacian in
directions $ \overrightarrow{z}$. The physical interpretation of
the function $K_1\left( x_i,z,t\right) $ is the bulk to boundary
propagator of massless vector field in the $d+1$-dimensional $AdS$
spacetime with the radial coordinate $t=z_0^2$. Now we can write
(\ref{48}) in terms of this propagator in a more suitable form to
match with the two-point supergravity correlator (\ref{22}): \bea
\label{54} \Pi^{ab}_{{\rm scalar}} \left( x_1,x_2\right)
&=&\int\limits_0^\infty \frac{dt}{t^{d/2+1}}\int\limits_{-\infty
}^\infty d^dz\ K_1\left(
x_1;z,t\right) K_1\left( x_2;z,t\right)  \nn\\
&& \times  \left[ -\frac{C^{ab}_1}{\Gamma
\left(\frac{d}{2}+1\right) }\epsilon _1\cdot \epsilon _2t+\frac
{C^{ab}_2}{\Gamma \left(\frac{d}{2}\right) }\epsilon _1\cdot
\left( z-x_2\right) \epsilon _2\cdot \left( z-x_1\right) \right].
\eea Comparing (\ref{54}) with the two-point correlation function
$\Gamma \left( x_1,x_2\right) $ for free scalar field theory, we
find the additional square bracket factor in our case, which
should be replaced by $t^3$ for free scalar theory.  Thus, the
gluon polarization operator is expressed in terms of bulk to
boundary propagator $K_1\left( x;z,t\right) $ of a massless field.

For vector-spinor interaction case we have spinor particles in the
loop and the two-point function is obtained from the one loop
spinor effective action. Setting $N=2$ in (\ref{30}) and taking
integrals over the Grassmann variables, it gets the following form
\cite{14}:
\begin{equation}
\label{56} \Pi^{ab}_{{\rm spinor}} \left( k_1,k_2\right) =
2Tr\left(T^aT^b\right)\frac{\left( g\mu ^{\epsilon/2}\right)
^2}{\left( 4\pi \right) ^{2-\epsilon/2}}\int\limits_0^\infty
\frac{d\tau }{\tau ^{d/2-1}} \int\limits_0^1d\alpha \ e^{\tau
k_1k_2\alpha \left( 1-\alpha \right) } \int\limits_{-\infty
}^\infty \frac{d^dz}{\left( 2\pi
\right) ^d}e^{i\left( k_1+k_2\right) z} \nn\\
$$
$$
\times \left[ \frac 2\tau \left( \delta \left( \alpha \right)
-1\right) \epsilon _1\cdot \epsilon _2 +\left[ \left( 1-2\alpha
\right) ^2-1\right] \left( \epsilon _1\cdot k_2\right) \left(
\epsilon _2\cdot k_1\right) +\left( \epsilon _1\cdot \epsilon
_2\right) \left( k_1\cdot k_2\right) \right] .
\end{equation}
The worldline expression of two-point correlator for this
interaction has minor difference with the scalar loop case in
square bracket in (\ref{34}) \cite{15,17}. This enables us to
re-express the spinor loop case in (\ref{56}) similar to
(\ref{54}) as:
\begin{equation}
\label{58} \Pi^{ab}_{{\rm spinor}} \left( x_1,x_2\right)
=2\int\limits_0^\infty \frac{dt}{t^{d/2+1}}\int\limits_{-\infty
}^\infty d^dz\ K_1\left( x_1;z,t\right) K_1\left( x_2;z,t\right)
\left[ C^{ab}_1\frac{1}{\Gamma \left(d/2+1\right) }\epsilon
_1\cdot
\epsilon _2t\right. +\nn\\
$$
$$
\left.+\frac1{\Gamma \left(d/2\right) }\ \left[ C^{ab}_2\epsilon
_1\cdot \left( z-x_2\right) \epsilon _2\cdot \left( z-x_1\right)
-C^{ab}_3\epsilon _1\cdot \epsilon _2\left( z-x_1\right) \cdot
\left( z-x_2\right) \right] \right] .
\end{equation}
$$
C^{ab}_3=Tr\left(T^aT^b\right)C^{\prime }\int\limits_0^1d\alpha \
\left[ \alpha \left( 1-\alpha \right) \right] ^{d/2-2}
$$

Thus, we see from (\ref{54}) and (\ref{58}) that in $AdS$ space
the two-point correlators of massless vector field interacted with
the scalar and spinor fields are expressed by means of massless
bulk to boundary propagator in this spacetime.

Before adding the contributions of scalar and fermion loops, we
should recall that our fermions are in ${\bf 4}$ and scalars are
in ${\bf 6}$ representations of $SU(4)$ group, they have $1/2$ and
$1$ quadratic casimir, respectively. So we add 2 times of scalar
loop contribution to the fermion one to find:
$$
\Pi ^{ab}\left( x_1,x_2\right) =4C_2\frac 1{\Gamma \left(
d/2\right) }\epsilon _1^i\epsilon _2^j\int\limits_0^\infty
\frac{dt}{t^{d/2+1}}\int d^dz\ K_1\left( x_1\right) K_1\left(
x_2\right)
$$
\begin{equation}
\label{59}\times \left[ \left( z-x_1\right) _i\left( z-x_2\right)
_j-\delta _{ij}\left( z-x_1\right) \cdot \left( z-x_2\right)
\right].
\end{equation}

As we will see later, the above result can be interpreted as a
correlation obtained in the bulk theory, where $K_1$ functions are
the bulk to boundary propagators and the square bracket show the
tensorial structure. We will be back on this in section 4.

%%%%%%%%%%%%%%%%%%%%%%%%%%%%%%%%%%%%%%%%%%%%%%%%%%%%%%%%%%%%%%%%%%%%%%%%%%%%%%%%%%
\Section{The three-point function}

At one loop approximation the field theory three-point correlation
function is the loop (scalar or spinor) having three external
vector field lines. Physically it describes photon or gluon
splitting amplitude due to the Dirac vacuum of spinor (scalar)
particles. According to Furry's theorem in QED an amplitude with
an odd number of external vector lines is zero in absence of
background fields \cite{18,15}. In non-abelean theory it becomes
non-zero due to color matrices in the internal lines. The starting
expression for scalar loop contribution to three-point correlation
function of non-abelean vector field can be found from (\ref{28})
by setting $N=3$, and changing variables
$\alpha_i=\left|u_i-u_{i+1}\right|$:
\begin{equation}
\label{86} \Gamma^{abc} \left( k_1,k_2,k_3\right)
=Tr\left(T^aT^bT^c\right) \frac{\left( ig\mu ^{\epsilon/ 2}\right)
^3}{\left( 4\pi \right) ^{2-\epsilon/2}}\delta ^{\left( d\right)
}\left( \sum k_l\right)\int\limits_0^\infty \frac{d\tau }{\tau ^{d/2-2}}%
\prod\limits_{l=1}^3\int\limits_0^1d\alpha _l\delta \left( \sum
\alpha _l-1\right)
$$
$$ \times \ e^{-\tau\left(k_1^2\alpha _2\alpha _3+k_2^2\alpha _1\alpha
_3+k_3^2\alpha _1\alpha_2\right) }\left[ \exp
\sum\limits_{i<j}\left[
-i\left( k_i\cdot \epsilon _j-k_j\cdot \epsilon _i\right) \stackrel{.}{G}%
_B^{ji}+\epsilon _i\cdot \epsilon _j\stackrel{..}{G}_B^{ji}\right]
\right] \mid _{linear\ in\ each\ \epsilon },
\end{equation}
Here in the first exponent we have used $k_i\cdot
k_j=\frac12\left({k_k}^2-{k_i}^2-{k_j}^2\right)$, that is the
result of the energy-momentum conservation. The above amplitude is
distinct from the three-point correlation function for the free
scalar field theory by an additional square bracket factor
\cite{16}. Decomposing this square bracket and keeping the terms
which have only first degree of each $\epsilon _i$ $,$ we get the
following expression for this bracket\footnote{We drop out
$\epsilon^{ijk}$, keeping in mind that indices $ i,j,k $ get
different values.}:
$$
-i\sum\limits_{i<j}\left( \epsilon _i\cdot \epsilon _j
\stackrel{..}{G}_B^{ji}+k_i\cdot \epsilon _jk_j\cdot \epsilon
_i\left(
\stackrel{.}{G}_B^{ji}\right) ^2\right) \left( k_i\cdot \epsilon _k\stackrel{%
.}{G}_B^{ki}+k_j\cdot \epsilon _k\stackrel{.}{G}_B^{kj}\right)  \\
$$
\begin{equation}\label{88}
-i\sum\limits_{i<j,k} k_i\cdot \epsilon _j\ k_j\cdot
\epsilon _k k_k\cdot \epsilon _i \stackrel{.}{G}_B^{ij}%
\stackrel{.}{G}_B^{jk}\stackrel{.}{G}_B^{ki}
\end{equation}

Going to the position space by Fourier transformation we integrate
over the momenta and write the result in the heat kernel terms
using (\ref{40}):
$$
\int dk_i\ \exp \left\{ -\tau k_i^2\alpha _j\alpha _k+ik_i\left(
z-x_i\right) \right\} =
$$
$$
\left( \frac \pi {\tau \alpha _j\alpha _k}\right)
^{1/2}\exp \left\{ -\frac{\left( z-x_i\right) ^2}{4\tau \alpha _j\alpha _k}%
\right\} =\left( 2\pi \right) ^d\left\langle z\left| e^{\tau
\alpha _j\alpha_k\Box }\right| x_i\right\rangle.
$$
In square bracket factor we replace momenta as follows:
$k_j\rightarrow\left( i\frac
\partial {\partial
x_j}\right)\rightarrow-i\frac{\left(x_j-z\right)}{2\tau\alpha_i\alpha_k}$.
Integrals over the $\alpha_i$ in position space will have the
form:
\begin{equation}\label{90}
\int\limits_0^\infty d\tau\frac 1{
\tau^{d/2-2}}\int\limits_0^1d\alpha _1\int\limits_0^1d\alpha
_2\int\limits_0^1d\alpha _3\delta \left( \sum \alpha _i-1\right)
\int d^dz\prod\limits_{i=1}^3\left\langle x_i\left| e^{\tau\alpha
_j\alpha _k\Box}\right| z\right\rangle .
\end{equation}
For three-point function the change of variables is the following
one \cite{16}:
$$
t=4\tau\alpha _1\alpha _2\alpha _3\rho,\ \tau \alpha
_j\alpha_k=\frac t{4 \rho _i},\ \rho _i=\rho\alpha _i.
$$
Then the integral (\ref{90}) changes to:
$$
4^{d/2-3}\int\limits_0^\infty \frac{dt}{t^{d/2-2}}\ \rho
^{-d+4}\prod\limits_{i=1}^3\int\limits_0^\rho d\rho _i\ \rho
_i^{d/2-3}\delta \left( \sum \rho _i-\rho\right) \int
d^dz\left\langle x_i\left| e^{\frac t{4\rho _i}\Box }\right|
z\right\rangle
$$
In these variables $\stackrel{.}{G}_B^{ij}$,
$\stackrel{..}{G}_B^{ij}$ and momenta can be written as:
$$
\stackrel{.}{G}_B^{ij}=\left( -1\right) ^{F_{ij}}\left( 1-2\rho
^{-1}\rho
_k\right) ,\ \stackrel{..}{G}_B^{ij}=\frac{8\rho ^{-2}}t\prod\limits_{l=1}^3%
\rho _l\left( \rho \delta \left( \rho _k\right) -1\right),\
k_i\rightarrow -i\frac{2\rho _i}t\left( z-x_i\right).
$$
Taking these expressions into account in (\ref{88}), the
additional square bracket factor has the following explicit form
in the new variables $\rho _i$ and $\ t$:

$$
-\frac 8{t^2}\prod\limits_{l=1}^3\rho _l\sum\limits_{i<j}\left[
2\rho ^{-2}\epsilon _i\cdot \epsilon _j+\frac 1t\epsilon _i\cdot
\left( z-x_j\right) \epsilon _j\cdot \left( z-x_i\right) \left(
\rho _k^{-1}-4\rho ^{-1}+4\rho ^{-2}\rho _k\right) \right]
$$
$$
\times \left[ \left( -1\right) ^{F_{ik}}\epsilon _k\cdot \left(
z-x_i\right) \rho _i\left( 1-2\rho ^{-1}\rho _j\right) +\left(
-1\right) ^{F_{jk}}\epsilon _k\cdot \left( z-x_j\right) \rho
_j\left( 1-2\rho ^{-1}\rho _i\right) \right]
$$
$$
-\frac 8{t^3}\prod\limits_{l=1}^3\rho _l\sum\limits_{i<j,k}\left(
-1\right) ^{F_{ij}+F_{ik}+F_{jk}}\epsilon _i\cdot \left(
z-x_k\right) \epsilon _k\cdot \left( z-x_j\right) \epsilon _j\cdot
\left( z-x_i\right)
$$
\begin{equation}
\label{92}\times \left[ 1-4\rho ^{-2}\left( \rho _i\rho _j+\rho
_i\rho _k+\rho _k\rho _j\right) +8\rho ^{-3}\rho _i\rho _j\rho
_k\right] .
\end{equation}

Notice that in (\ref{92}) for three-point function we cannot
factor out $\alpha$ -dependent integrals and introduce an integral
over $\beta$, as we made for two-point function case, since here
we have different degrees of $\rho _i$ and $\rho$. This means that
we should introduce different  $\Gamma \left( d-m\right) $ and
$K_n\left( x,z,t\right) $ functions corresponding to these degrees
of $\rho$ and $\rho _i$\footnote{Mathematically correct way is to
introduce $\Gamma$-functions first , then to make a change of
variables as we did for two-point function.}. So, the three-point
function, in addition of $K_1$, includes $K_2$ and $K_3$ functions
which are defined as:
\begin{equation}
\label{94}
\begin{array}{c}
K_n\left( x_i,z,t\right) \equiv \int\limits_0^\infty d\rho _i\rho
_i^{d/2-n}e^{-\rho _i}\left\langle x_i\left| e^{\frac t{4\rho
_i}\Box }\right| z\right\rangle .
\end{array}
\end{equation}
The appearance of these functions in the three-point function is
the reflection of derivative interactions in the bulk theory.
Indeed,
$$\frac{\p K_n}{\p x^\mu} = -\frac{2}{t} (x^\mu-z^\mu)K_{n-1} $$

Thereby, rewritten in terms of $K_n\left( x_i\right)$ functions
the three-point correlation function (\ref{86}) will have the
following form:
$$
\Gamma ^{abc}\left( x_1,x_2,x_3\right) =\frac{i2^{d/2-3}\left(
g\mu ^{\epsilon /2}\right) ^3}{\left( 2\pi \right) ^{3/2}}Tr\left(
T^aT^bT^c\right) \int\limits_0^\infty
\frac{dt}{t^{d/2+1}}\int\limits_{-\infty }^\infty d^dz\ \left[
\sum\limits_{i<j}\left\{ \left( z-x_i\right) \cdot \epsilon
_k\times \right. \right.
$$
$$
\left[ \left( -1\right) ^{F_{ik}}\cdot K_1\left( x_i\right)
K_2\left( x_j\right) +\left( -1\right) ^{F_{jk}}K_1\left(
x_j\right) K_2\left( x_i\right) \right] \left[ \frac 2{\Gamma
\left( d-1\right) }t\epsilon _i\cdot \epsilon _jK_2\left(
x_k\right) +\right.
$$
$$
\left. \left( z-x_i\right) \cdot \epsilon _j\left( z-x_j\right)
\cdot \epsilon _i\left( \frac 1{\Gamma \left( d-3\right)
}K_3\left( x_k\right) -\frac 4{\Gamma \left( d-2\right) }K_2\left(
x_k\right) +\frac 4{\Gamma \left( d-1\right) }K_1\left( x_k\right)
\right) \right]
$$
$$
-2\left( \left( -1\right) ^{F_{ik}}+\left( -1\right)
^{F_{jk}}\right) \left( z-x_i\right) \cdot \epsilon _k\left[ \frac
2{\Gamma \left( d\right) }t\epsilon _i\cdot \epsilon _jK_2\left(
x_k\right) +\right.
$$
$$
\left. \left. \left( z-x_i\right) \cdot \epsilon _j\left(
z-x_j\right) \cdot \epsilon _i\left( \frac 1{\Gamma \left(
d-2\right) }K_3\left( x_k\right) -\frac 4{\Gamma \left( d-1\right)
}K_2\left( x_k\right) +\frac 4{\Gamma \left( d\right) }K_1\left(
x_k\right) \right) \right] \right\}
$$
$$
+\sum\limits_{i<j,k}\left( -1\right) ^{F_{jk}}\left( z-x_i\right)
\cdot \epsilon _j\left( z-x_j\right) \cdot \epsilon _k\left(
z-x_k\right) \cdot \epsilon _i\left\{ \frac 1{\Gamma \left(
d-3\right) }K_2\left( x_i\right) K_2\left( x_j\right) K_2\left(
x_k\right) \right.
$$
$$
-\frac 4{\Gamma \left( d-1\right) }\left[ K_2\left( x_i\right)
K_1\left( x_j\right) K_1\left( x_k\right) +K_1\left( x_i\right)
K_2\left( x_j\right) K_1\left( x_k\right) +K_1\left( x_i\right)
K_1\left( x_j\right) K_2\left( x_k\right) \right]
$$
\begin{equation}
\label{96}\left. \left. +\frac 8{\Gamma \left( d\right) }K_1\left(
x_i\right) K_1\left( x_j\right) K_1\left( x_k\right) \right\}
\right]
\end{equation}
As is seen here, the three-point correlation function is expressed
by means of bulk to boundary propagators of massless fields and
its descendants.  This is in contrast to the free scalar field
theory where the three-point correlation function was expressed in
terms of a unique bulk to boundary propagator of a massive scalar
field \cite{16}. From the $AdS/CFT$ correspondence dictionary, we
expect the appearance of the massless propagator, $K_1$, as it
happened in the two-point function. Indeed, $K_2$ and $K_3$
functions are not really independent propagators and their
appearance, as mentioned before, is due to derivative
interactions.

The fermionic loop contribution to the three-point function of
non-abelean vector field is not zero. This is in contrast to the
abelean theory, where contribution of one-loop diagrams with odd
number vertices vanishes due to Furry's theorem. The contribution
of this loop can be obtained from formula (\ref{30}) by setting
$N=3$ . Decomposing the exponent, keeping only the terms linear in
each $\epsilon$ and then taking integrals over the Grassmann
variables we obtain:
$$
\Gamma^{abc} \left( k_1,k_2,k_3\right) =-2\frac{\left( ig\right)
^3}{\left( 4\pi \right) ^2}Tr\left( T^aT^bT^c\right)
\int\limits_0^\infty d\tau \
\prod\limits_{l=1}^3\int\limits_0^1d\alpha _l\ e^{-\tau \left(
k_1^2\alpha _2\alpha _3+k_2^2\alpha _1\alpha _3+k_3^2\alpha
_1\alpha _2\right) }
$$
$$
\times \delta \left( \sum k_i\right)\delta \left( \sum \alpha
_i-1\right)\left(-i\right) \left[ {\rm scalar\ loop\ terms\ }
+\left(k_1\cdot k_2\ \epsilon_1\cdot \epsilon_2-k_1\cdot \epsilon
_2\ \ k_2\cdot \epsilon _1\right)\times \right.
$$
$$
\left( k_1\cdot \epsilon _3%
\stackrel{.}{G}_B^{31}+k_2\cdot \epsilon _3\stackrel{.}{G}_B^{32}\
\right)+\left(k_1\cdot k_3\ \epsilon_1\cdot \epsilon_3-k_1\cdot
\epsilon _3\ \
k_3\cdot \epsilon _1\right)\left( k_1\cdot \epsilon _2%
\stackrel{.}{G}_B^{21}-k_3\cdot \epsilon _2\stackrel{.}{G}_B^{32}\
\right)
$$
$$
-\left(k_2\cdot k_3\ \epsilon_2\cdot \epsilon_3 -k_2\cdot \epsilon
_3\ k_3\cdot
\epsilon _2\right)\left( k_2\cdot \epsilon _1%
\stackrel{.}{G}_B^{21}+k_3\cdot \epsilon _1\stackrel{.}{G}_B^{31}\
\right)+\left[\left(k_1\cdot \epsilon_2\ \epsilon_1\cdot
\epsilon_3-\epsilon_1\cdot \epsilon_2\ \ k_1\cdot
\epsilon_3\right) \right.
$$
$$
\times k_2\cdot k_3-\left(\epsilon_1\cdot k_2\ \ k_1\cdot k_3-
k_1\cdot k_2\ \epsilon_1\cdot k_3\right)\epsilon_2\cdot
\epsilon_3+\left(\epsilon_1\cdot \epsilon_2\ \ k_1\cdot
k_3-k_1\cdot \epsilon _2\ \ k_3\cdot \epsilon _1\right)k_2\cdot
\epsilon _3
$$
\begin{equation}
\label{98}\left.\left.  -\left(k_1\cdot k_2\ \epsilon_1\cdot
\epsilon_3- k_1\cdot \epsilon _3\ \ k_2\cdot
\epsilon_1\right)k_3\cdot \epsilon
_2\right]G_F^{21}G_F^{31}G_F^{32}\right]
\end{equation}
Since our fermions are in ${\bf 4}$ and scalars are in ${\bf 6}$
representations of $SU(4)$ group, they have $1/2$ and $1$
quadratic casimir, respectively. So we should add 2 times of
scalar loop contribution (\ref{86}) to the fermionic loop
contribution (\ref{98}). Scalar loop terms again are canceled and
in the square bracket only the following terms remain:
$$
\left[ \left(k_1\cdot k_2\ \epsilon_1\cdot \epsilon_2-k_1\cdot
\epsilon _2\ \
k_2\cdot \epsilon _1\right)\left( k_1\cdot \epsilon _3%
\stackrel{.}{G}_B^{31}+k_2\cdot \epsilon _3\stackrel{.}{G}_B^{32}\
\right) +\left(k_1\cdot k_3\ \epsilon_1\cdot \epsilon_3-k_1\cdot
\epsilon _3\ \ k_3\cdot \epsilon _1\right) \right.
$$
$$
\times\left( k_1\cdot \epsilon _2%
\stackrel{.}{G}_B^{21}-k_3\cdot \epsilon _2\stackrel{.}{G}_B^{32}\
\right)-\left(k_2\cdot k_3\ \epsilon_2\cdot \epsilon_3 -k_2\cdot
\epsilon _3\
k_3\cdot \epsilon _2\right)\left( k_2\cdot \epsilon _1%
\stackrel{.}{G}_B^{21}+k_3\cdot \epsilon _1\stackrel{.}{G}_B^{31}\
\right)
$$
$$
-\left[\left(k_1\cdot \epsilon_2\ \epsilon_1\cdot
\epsilon_3-\epsilon_1\cdot \epsilon_2\ \ k_1\cdot
\epsilon_3\right)k_2\cdot k_3-\left(\epsilon_1\cdot k_2\ \
k_1\cdot k_3- k_1\cdot k_2\ \epsilon_1\cdot
k_3\right)\epsilon_2\cdot \epsilon_3+ \right.
$$
\begin{equation}
\label{100}\left.\left.+\left(\epsilon_1\cdot \epsilon_2\ \
k_1\cdot k_3-k_1\cdot \epsilon _2\ \ k_3\cdot \epsilon
_1\right)k_2\cdot \epsilon _3 -\left(k_1\cdot k_2\ \epsilon_1\cdot
\epsilon_3- k_1\cdot \epsilon _3\ \ k_2\cdot
\epsilon_1\right)k_3\cdot \epsilon _2 \right]\right].
\end{equation}
In position space with $d=4$ by repeating the procedure of change
of variables as we did in scalar loop case, we get the expression
of three-point function in terms of $K_n (x_i)$ functions:
$$
\Gamma _3^{abc}\left( x_1,x_2,x_3\right) =\frac{i\left( g \right)
^3}{\left( 2\pi \right) ^{3/2}}Tr\left(
T^aT^bT^c\right)\int\limits_0^\infty \frac{dt}{t^3}\int d^4z\
\times
$$
$$
\left\{\left[\epsilon_1\cdot
\epsilon_2\left(z-x_1\right)\cdot\left(z-x_2\right) -\left(
z-x_1\right) \cdot \epsilon _2\left( z-x_2\right) \cdot \epsilon
_1\right]K_3\left( x_3\right) \left[ \left( z-x_1\right) \cdot
\epsilon _3\ \times\right. \right.
$$
$$
\left.K_1\left( x_1\right) \left( K_2\left( x_2\right) -K_1\left(
x_2\right) \right) + \left( z-x_2\right) \cdot \epsilon
_3K_1\left( x_2\right) \left( K_2\left( x_1\right) -K_1\left(
x_1\right) \right) \right]
$$
$$
\left.+\left[
\epsilon_1\cdot\epsilon_3\left(z-x_1\right)\cdot\left(z-x_3\right)-\left(z-x_1\right)
\cdot \epsilon _3\ \left( z-x_3\right) \cdot \epsilon _1\right]
K_3\left( x_2\right)\times \right.
$$
$$
\left[ \left( z-x_1\right) \cdot \epsilon _2\ K_1\left( x_1\right)
\left( K_2\left( x_3\right) -K_1\left( x_3\right) \right) -\left(
z-x_3\right) \cdot \epsilon _2K_1\left( x_3\right) \ \left(
K_2\left( x_1\right) -K_1\left( x_1\right) \right) \right] \
$$
$$
-\left[\epsilon_2\cdot\epsilon_3\left(z-x_2\right)\cdot\left(z-x_3\right)-\left(
z-x_2\right) \cdot \epsilon _3\ \left( z-x_3\right) \cdot \epsilon
_2\right]K_3\left( x_1\right) \left[ \left( z-x_2\right) \cdot
\epsilon _1K_1\left( x_2\right)\times \right.
$$
$$
\left.\left( K_2\left( x_3\right) -K_1\left( x_3\right)
\right)+\left( z-x_3\right) \cdot \epsilon _1K_1\left( x_3\right)
\left( K_2\left( x_2\right) -K_1\left( x_2\right) \right) \right]
$$
$$
 +K_2\left(
x_1\right) K_2\left( x_2\right) K_2\left(
x_3\right)\left[\left[\epsilon_1\cdot
\epsilon_3\left(z-x_1\right)\cdot \epsilon_2-\epsilon_1\cdot
\epsilon_2\left(z-x_1\right)\cdot
\epsilon_3\right]\left(z-x_2\right)\cdot \left(z-x_3\right)-
\right.
$$
$$
-\left[\epsilon_1\cdot \left(z-x_2\right)\left(z-x_1\right)\cdot
\left(z-x_3\right)- \left(z-x_1\right)\cdot
\left(z-x_2\right)\epsilon_1\cdot
\left(z-x_3\right)\right]\epsilon_2\cdot \epsilon_3
$$
$$
+\left[\epsilon_1\cdot \epsilon_2\left(z-x_1\right)\cdot
\left(z-x_3\right)-\left(z-x_1\right)\cdot
\epsilon_2\left(z-x_3\right)\cdot
\epsilon_1\right]\left(z-x_2\right)\cdot \epsilon_3
$$
\begin{equation}
\label{102} \left.\left. -\left[\epsilon_1\cdot
\epsilon_3\left(z-x_1\right)\cdot
\left(z-x_2\right)-\left(z-x_1\right)\cdot
\epsilon_3\left(z-x_2\right)\cdot
\epsilon_1\right]\left(z-x_3\right)\cdot \epsilon_2\right]
\right\}
\end{equation}

This result is considered as a realization for the low energy
limit of the open string calculation of R-current correlations.
According to the open/closed string duality, we expect that this
could be translated to the $AdS$ amplitude. Introducing the $AdS$
set-up and comparing the results in  the bulk and boundary
theories are the subjects of the following sections.

%%%%%%%%%%%%%%%%%%%%%%%%%%%%%%%%%%%%%%%%%%%%%%%%%%%%%%%%%%%%%%%%%%%%%%%%%%%%%%%%
\Section{The Set-up in $AdS$ space}

Let us introduce the $AdS$ supergravity fields which are the
corresponding partners of SYM R-currents. These $AdS$ fields are
$SU(4)$ gauge fields with the following action \cite{21}:
\bea\label{10} S[A]&=& \frac{1}{2g^2} \int \delta_{ab} dA^a \wedge
*dA^b + f_{abc}dA^a \wedge
*\{A^b\wedge A^c\} \nn\\
&&+\frac{ik}{32\pi^2}\int d_{abc} A^a \wedge dA^b \wedge dA^c \; ,
\eea where integrals are over 5-dimensional $AdS$-space and the
last term is the Chern-Simons term.

The suitable coordinate for our purposes is the Poincare
coordinate in Euclidean signature: \bea\label{12} ds^2=
\frac{1}{z_0^2} (d \vec{z}^{\; 2}+dz_0^2)\;,\eea where the
boundary of $AdS$ is at $z_0=0$.

Now we are ready to write the solutions of the classical equations
of motion. They would be determined in terms of their boundary
values, $a_i^a(\vec{x})$ as follows \cite{21}: \bea\label{14}
A^a(z_0,\vec{z})&=&dz_i \int d^dx
a_i^a(\vec{x})\frac{z_0^{d-2}}{[z_0^2+\left|\vec{z}-\vec{x}\right|^2]^{d-1}}
\nn\\
&&-dz_0 \int d^dx a_i^a(\vec{x})\frac{1}{2(d-2)}\frac{\p}{\p
x^i}\frac{z_0^{d-3}}{[z_0^2+\left|\vec{z}-\vec{x}\right|^2]^{d-2}}
\;.
\eea Plugging this solution in the action will give us  the
effective action in tree level as a functional of boundary values
\cite{21}: \bea
S[a]&=&\frac{1}{2g^2} \int d^d x_1 d^d x_2 a_i^a(x_1)a_j^b(x_2) \delta_{ab} \nn\\
&& \times
\left\{(\p_1^i\p_1^j-\delta^{ij}\p_1^2)I_{d-2,d-2}^{d/2-1} -
\frac{1}{8(d-2)^2} (\p_1^i\p_1^j-\delta^{ij}\p_1^2)\p_1^2
I_{d-3,d-3}^{d/2-2}
\right\} \nn\\
&&+ \frac{1}{2}g^2 \int d^d x_1 d^d x_2 d^d x_3
a_i^a(x_1)a_j^b(x_2)a_k^c(x_3)
f_{abc} \nn\\
&& \times \left\{2 \delta^{ij}\p_{1k} I_{d-2,d-2,d-2}^{d-2} +
\frac{1}{2(d-2)^2}
\p_{2j}(\p_1^i\p_1^k-\delta^{ik}\p_1^2) I_{d-3,d-3,d-2}^{d-3} \right\} \nn\\
&&+ \frac{ik}{32\pi^2} \int d^4 x_1 d^4 x_2 d^4 x_3
a_i^a(x_1)a_j^b(x_2)a_k^c(x_3)
d_{abc} \epsilon^{jklm} \frac{1}{d-2} \nn\\
&&\label{16} \times \left\{-\frac{1}{2} \p_{1i}\p_{2m}\p_{3l}
I_{d-3,d-2,d-2}^{3d/2-4} +
\p_{3l}(\p_{1i}\p_{1m}-\delta^{im}\p_1^2) I_{d-3,d-2,d-2}^{3d/2-4}
\right\}, \eea with \bea
 I_{mn}^f(x_1,x_2)&\equiv& \int dz_0 d^dz
\frac{z_0^{2f+1}}{[z_0^2+\left|\vec{z}-\vec{x_1}\right|^2]^{m+1}
[z_0^2+\left|\vec{z}-\vec{x_2}\right|^2]^{n+1}}\nn\\
\label{18} &=& \frac{\pi^d}{2\; m!\; n!} \int \frac{dt\;
d^dz}{t^{(m+n-f-d+2)}} K_{d-m}(x_1)K_{d-n}(x_2)  \;,  \\
 I_{mnp}^f(x_1,x_2,x_3)&\equiv& \int dz_0 d^dz
\frac{z_0^{2f+1}}{[z_0^2+\left|\vec{z}-\vec{x_1}\right|^2]^{m+1}
[z_0^2+\left|\vec{z}-\vec{x_2}\right|^2]^{n+1}[z_0^2+\left|\vec{z}-\vec{x_3}\right|^2]^{p+1}}
\nn\\
\label{20} &=& \frac{\pi^{3d/2}}{2\; m!\; n!\; p!} \int \frac{dt\;
d^dz}{ t^{(m+n+p-f-3d/2+3)}} K_{d-m}(x_1)K_{d-n}(x_2) K_{d-p}(x_3)
\;,\eea where we have used the following expression for $K_n$
functions with $t=z_0^2$,
$$ K_n\left( x;z,t\right) \equiv \int\limits_0^\infty d\rho \rho
^{d/2-n}e^{-\rho }\left\langle x\left| e^{\frac{t}{4\rho }\Box
}\right| z\right\rangle = \frac{(d-n)!}{(\pi t)^{d/2}}
\left(\frac{t}{t+|\vec{z}-\vec{x}|^2}\right)^{d-n+1}.
$$

Note that the solution (\ref{16}) is in zeroth order in the
Yang-Mills coupling and there is no need for ghosts, since we are
working in the tree level. The next step would be introducing this
effective action as generating functional of SYM boundary
operators.  So we will find two-point and three-point functions of
the R-currents by taking functional derivatives from this
generating function with respect to boundary fields
$a_i^a(\vec{x})$. The two-point function will be found as
\cite{21}: \bea\label{22} \left\langle
J_a^i(x_1)J_b^j(x_2)\right\rangle = -\frac{1}{g^2} \delta_{ab}
(\p_1^i\p_1^j-\delta^{ij}\p_1^2) I_{22}^1 \; . \eea The
three-point function includes two parts coming from $f_{abc}$ and
$d_{abc}$ parts of the effective action: \bea\label{24}
&&\left\langle J_a^i(x_1)J_b^j(x_2)J_c^k(x_3)\right\rangle_{f_{abc}} \nn \\
 &&= -\frac{f_{abc}}{g^2}\left\{ \delta_{ij}(\p_{1k}-\p_{2k})I^2_{222} +\frac{1}{16}
\left[ \p_{2j}(\p_{1i}\p_{1k}-\delta_{ik}\p_1^2)\right] I^1_{112}
+{\rm perm.} \right\} \;,\eea \bea\label{26}
&&\left\langle J_a^i(x_1)J_b^j(x_2)J_c^k(x_3)\right\rangle_{d_{abc}} \nn \\
 &&= -\frac{ikd_{abc}}{64\pi^2}\left\{ \epsilon_{jklm} \left[
(\p_{1i}\p_{1l}-\delta_{il}\p_1^2)(\p_{2m}-\p_{3m}) +
\p_{1i}\p_{2l}\p_{3m} \right] I^2_{122}  +{\rm perm.} \right\}\;,
\eea where `perm.' stands for permutating indices $\{i,j,k\}$ and
$\{1,2,3\}$ to find a totally (anti)symmetric expression for
$(f)d_{abc}$ part. These permutations should also be carried on
lower indices of $I_{112}^1$ and $I_{122}^2$.

%%%%%%%%%%%%%%%%%%%%%%%%%%%%%%%%%%%%%%%%%%%%%%%%%%%%%%%%%%%%%%%%
\begin{figure}\label{fig3}
 \begin{center}
  \includegraphics[scale=.6]{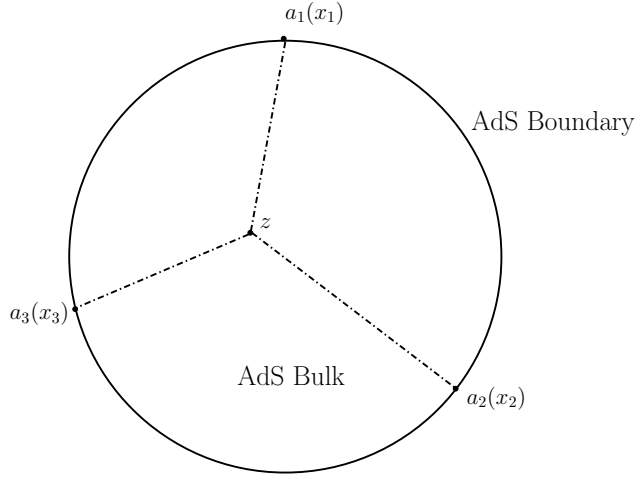}
 \end{center}
\caption{The three-point function in the $AdS$ space. The
dotted-dashed lines are bulk to boundary propagators from the
point $z$ to points $x_i$'s on the boundary. $a_i$'s are the
boundary values for the vector fields which are sources for
R-currents in the SYM theory. }
\end{figure}

%%%%%%%%%%%%%%%%%%%%%%%%%%%%%%%%%%%%%%%%%%%%%%%%%%%%%%%%%%%%

Now inserting the expressions (\ref{18}) and (\ref{20}) in the
above two- and three-point functions, we find these correlations
in terms of $K_n$ functions:
$$
\left\langle J_a^i\left( x_1\right) J_b^j\left( x_2\right)
\right\rangle =\frac 1{4g^2}\delta _{ab}\int\limits_0^\infty
\frac{dt}{t^{d/2+1}} t^{2-d/2}\int d^dz\ K_1\left( x_1\right)
K_1\left( x_2\right)
$$
\begin{equation}
\label{104}\times \left[ \left( z-x_1\right) _i\left( z-x_2\right)
_j-\delta _{ij}\left( z-x_1\right) \cdot \left( z-x_2\right)
\right]
\end{equation}
and for the $f_{abc}$ part of the three-point function:
\begin{equation}
\begin{array}{c}
\label{108} \left\langle J_a^i\left( x_1\right) J_b^j\left(
x_2\right) J_c^k\left( x_3\right)\right\rangle =
\frac{ \pi^6}8 f_{abc}\int\limits_0^\infty \frac{dt}{t^3}%
\ \int d^4z\ \nn\\
\times \left\{t K_2\left( x_3\right) \delta _{ij}\left[ \left(
z-x_1\right) _kK_1\left( x_1\right) K_2\left( x_2\right) -\left(
z-x_2\right) _kK_1\left( x_2\right) K_1\left( x_2\right) \right]\right. \\
\left. +tK_2\left( x_1\right) \delta _{jk}\left[ \left(
z-x_2\right) _iK_1\left( x_2\right) K_2\left( x_3\right) -\left(
z-x_3\right) _iK_1\left(
x_3\right) K_2\left( x_2\right) \right]\right. \\
+tK_2\left( x_2\right) \delta _{ki}\left[ \left( z-x_3\right)
_jK_1\left( x_3\right) K_2\left( x_1\right) -\left( z-x_1\right)
_jK_1\left( x_1\right)
K_2\left( x_3\right) \right] + \\
+K_2\left( x_3\right) K_1\left( x_1\right) K_2\left( x_2\right)
\left( z-x_2\right) _j\left[ \left( z-x_1\right) _k\left(
z-x_1\right) _i+\delta
_{ki}\left( z-x_1\right) ^2\right] - \\
-K_2\left( x_3\right) K_1\left( x_2\right) K_2\left( x_1\right)
\left( z-x_1\right) _i\left[ \left( z-x_2\right) _j\left(
z-x_2\right) _k+\delta
_{kj}\left( z-x_2\right) ^2\right] + \\
+K_2\left( x_1\right) K_1\left( x_2\right) K_2\left( x_3\right)
\left( z-x_3\right) _k\left[ \left( z-x_2\right) _i\left(
z-x_2\right) _j+\delta
_{ij}\left( z-x_2\right) ^2\right] - \\
-K_2\left( x_1\right) K_1\left( x_3\right) K_2\left( x_2\right)
\left( z-x_2\right) _k\left[ \left( z-x_3\right) _i\left(
z-x_3\right) _j+\delta
_{ij}\left( z-x_3\right) ^2\right] + \\
+K_2\left( x_2\right) K_1\left( x_3\right) K_2\left( x_1\right)
\left( z-x_1\right) _k\left[ \left( z-x_3\right) _i\left(
z-x_3\right) _j+\delta
_{ij}\left( z-x_3\right) ^2\right] - \\
\left. -K_2\left( x_2\right) K_1\left( x_1\right) K_2\left(
x_3\right) \left( z-x_3\right) _k\left[ \left( z-x_1\right)
_i\left( z-x_1\right) _j+\delta _{ij}\left( z-x_1\right) ^2\right]
\right\} .
\end{array}
\end{equation}

These are correlation functions derived from the supergravity
theory in the bulk. In the string theory language, they correspond
to a tree level closed string amplitude. Notice in these
expressions, the supergravity correlators contain only $
K_1\left(x_i\right)$ and $K_2\left(x_i\right)$ functions and
apparently differs from the super Yang-Mills correlator
(\ref{102}), which additionally contains $K_3 (x_i)$. However, as
mentioned before, $K_n\left(x_i\right)$ functions are related to
each other and it would be possible to bring these apparently
different expressions into a unique form. To avoid complications
involved here, we use an indirect way and show the equivalence of
worldline expression\footnote{This equivalence for two-point
function was shown in \cite{14}.} (\ref{98}) and its ordinary
expression in proper-time parameters $\tau _i$ given in \cite{21}.
Then compare it with the $AdS$ results. We will do this
manipulation in the following section.

%%%%%%%%%%%%%%%%%%%%%%%%%%%%%%%%%%%%%%%%%%%%%%%%%%%%%%%%%%%%%%%%%%%%%%%%%%%%%%
\Section{$AdS/CFT$ matching correlators}

Let us start with the scalar loop contribution in the SYM side.
Putting back the regrouped terms of (\ref{88}) into (\ref{86}) we
obtain the following form for it:

\bea\label{109}
\Gamma ^{abc} \left( k_1,k_2,k_3\right)  =Tr\left( T^aT^bT^c\right)  \frac{%
\left( ig\mu ^{\epsilon /2}\right) ^3}{\left( 4\pi \right) ^{2-\epsilon /2}}%
\delta ^{\left( d\right) }\left( \sum k_l\right)  \int\limits_0^\infty \frac{%
d\overline{\tau }}{\overline{\tau }^{d/2-2}}\prod\limits_{l=1}^3\int%
\limits_0^1d\alpha _l\delta \left( \sum \alpha _l-1\right)  \nn\\
i\left[
\overline{\tau }\left( \stackrel{.}{G}_B^{21}k_2\cdot \epsilon _1+\stackrel{.%
}{G}_B^{31}k_3\cdot \epsilon _1\right) \left(
\stackrel{.}{G}_B^{32}k_3\cdot \epsilon
_2-\stackrel{.}{G}_B^{21}k_1\cdot \epsilon _2\right) \left(
\stackrel{.}{G}_B^{31}k_1\cdot \epsilon
_3+\stackrel{.}{G}_B^{32}k_2\cdot \epsilon _3\right)\right. \nn\\
\left. + \sum\limits_{i<j} \epsilon _i\cdot \epsilon _j
\left( k_i\cdot \epsilon _k\stackrel{%
.}{G}_B^{ki}+k_j\cdot \epsilon
_k\stackrel{.}{G}_B^{kj}\right)\right]e^{-\overline{\tau }\left(
k_1^2\alpha _2\alpha _3+k_2^2\alpha _1\alpha _3+k_3^2\alpha
_1\alpha _2\right)} , \eea where we substituted
$\stackrel{..}{G}_B^{ij}$ omitting its $\delta \left( \alpha
_k\right) $ terms. As we know \cite{14} these terms in bosonic
worldline Green's function correspond to terms, which arise due to
tadpole diagrams of scalar field theory. To match worldline
correlators with the ones in \cite{21} we also ignore the
contribution of tadpole diagrams, and will be back to this point
in the end of this section.

Now plug the following relations in (\ref{109})
\begin{equation}
\label{114}\stackrel{.}{G}_B^{21}=\left( 1-2\alpha _3\right) ,\stackrel{.}{G}%
_B^{32}=\left( 1-2\alpha _1\right) ,\stackrel{.}{\
G}_B^{31}=-\left( 1-2\alpha _2\right) ,
\end{equation}
and make the change of variables $\tau _i=\overline{\tau }\alpha
_i$ in the integral. The Jacobian for the change $\left( \alpha
_1,\ \alpha _2,\ \overline{\tau }\right) \rightarrow \left( \tau
_1,\ \tau _2,\ \tau _3\right) $ is $J=1/\overline{\tau }^2$. Then
integrals in (\ref{109}) can be written as:

$$
\int\limits_0^\infty \overline{\tau }^2d\overline{\tau }\frac 1{\overline{%
\tau }^{d/2+1}}\int\limits_0^1d\alpha _1\int\limits_0^1d\alpha
_2\int\limits_0^1d\alpha _3\delta \left( \sum \alpha _l-1\right) \
e^{\sum \left( -\overline{\tau }\alpha _i\alpha _jk_k^2\right) }
$$
$$
=\int\limits_0^\infty d\tau _1d\tau _2d\tau _3\frac
1{\overline{\tau }^{d/2+1}}\ e^{-\left( \tau _1\tau _2\tau
_3\right) /\overline{\tau }\left( \sum k_i^2/\tau _i\right) }
$$
After these changes we can factor out $\epsilon_1^i\epsilon
_2^j\epsilon _3^k$ in (\ref{109}) and apply relations
$k_i+k_j+k_k=0$ and $\tau _1+\tau _2+\tau _3=\overline{\tau }$ to
convert brackets. In the result, (\ref{109}) can be written as a
cyclic sum of two terms:
$$
\Gamma ^{abc} \left( k_1,k_2,k_3\right)  =Tr\left( T^aT^bT^c\right)  \frac{%
\left( ig\mu ^{\epsilon /2}\right) ^3}{\left( 4\pi \right) ^{2-\epsilon /2}}%
\delta ^{\left( d\right) }\left( \sum k_l\right)
\int\limits_0^\infty
d\tau _1d\tau _2d\tau _3\frac{\pi ^{d/2}}{\overline{\tau }^{d/2+1}}e^{-\frac{%
\tau _1\tau _2\tau _3}{\overline{\tau }}\left( \sum k_i^2/\tau
_i\right)}
$$
$$
\times\epsilon_1^i\epsilon _2^j\epsilon _3^k \left[ -\frac
1{6\overline{\tau }^2}\left( 2\tau _1k_{2i}+\left( \overline{\tau
}-2\tau _2\right) k_{1i}\right) \left( 2\tau _2k_{3j}+\left(
\overline{\tau }-2\tau _3\right) k_{2j}\right) \left( 2\tau
_3k_{1k}+\left( \overline{\tau }-2\tau _1\right) k_{3k}\right)
\right.
$$
$$
+\frac 1{6\overline{\tau }^2}\left( 2\tau _1k_{3i}+\left( \overline{\tau }%
-2\tau _3\right) k_{1i}\right) \left( 2\tau _2k_{1j}+\left( \overline{\tau }%
-2\tau _1\right) k_{2j}\right) \left( 2\tau _3k_{2k}+\left( \overline{\tau }%
-2\tau _2\right) k_{3k}\right)
$$
$$
\left. +\frac 2{\overline{\tau }}\delta _{ij}\left( \tau _3\left(
k_{2k}-k_{1k}\right) +\left( \tau _1-\tau _2\right) k_{3k}\right)
+cyclic\right] .
$$

 We can consider the spinor loop contribution in
(\ref{98}) as well and do the above manipulation to find the
following expression as a combination of
scalar and spinor loops. Notice firstly, that in Yang-Mills correlator (\ref{102}%
) we have terms like $\left( z-x_i\right) \cdot \left(
z-x_j\right) $, which are absent in the supergravity one
(\ref{108}). This kind of terms comes from terms including
$k_i\cdot k_j$ in (\ref{98}). Since we are dealing with massless
vector fields $\left( k_l^2=0\right) $, we should put $k_i\cdot
k_j=0$ due to energy-momentum conservation law. As is seen from
(\ref{98}) the scalar loop terms will be cancelled with the terms
of scalar loop contribution, if we multiply the second one by a
factor of 2, as explained above (\ref{59}), and add it to
(\ref{98}). Now we reach to the following worldline expression for
this summation: \bea\label{117} &&\left( 2\pi \right)^{d/2}
\int\limits_0^\infty d\tau _1d\tau _2d\tau _3\frac 1{\overline{\tau }%
^{d/2+1}}\ e^{-\frac{\left( \tau _1\tau _2\tau
_3\right)}{\overline{\tau }\left( \sum k_i^2/\tau _i\right) }}
\left[ -k_1\cdot \epsilon _2\ k_2\cdot \epsilon
_1\left( k_1\cdot \epsilon _3\stackrel{.}{G}_B^{31}+k_2\cdot \epsilon _3%
\stackrel{.}{G}_B^{32}\ \right) \right. \nn\\
&&\left. -k_1\cdot \epsilon _3\ k_3\cdot \epsilon
_1\left( k_1\cdot \epsilon _2\stackrel{.}{G}_B^{21}-k_3\cdot \epsilon _2%
\stackrel{.}{G}_B^{32}\ \right)  +k_2\cdot \epsilon _3\
k_3\cdot \epsilon _2\left( k_2\cdot \epsilon _1\stackrel{.}{G}%
_B^{21}+k_3\cdot \epsilon _1\stackrel{.}{G}_B^{31}\ \right) \right.  \nn\\
&&\left. -k_1\cdot \epsilon _2\ k_2\cdot \epsilon _3\ k_3\cdot
\epsilon _1+k_1\cdot \epsilon _3\ k_3\cdot \epsilon _2\ k_2\cdot
\epsilon _1\right]. \eea

Inserting the expressions for $G_B$ factors and after some
algebra, we can write the vector field's three-point correlation
functions from the SYM theory which includes both the scalar and
spinor loop contributions as in the following: \bea\label{119}
\Gamma ^{abc}\left( k_1,k_2,k_3\right) = Tr\left( T^aT^bT^c\right) \frac{%
\left( ig\mu ^{\epsilon /2}\right) ^3}{\left( 4\pi \right) ^{2-\epsilon /2}}%
\delta ^{\left( d\right) }\left( \sum k_l\right)
\int\limits_0^\infty d\tau
_1d\tau _2d\tau _3\frac{\pi ^{d/2}}{\overline{\tau }^{d/2+1}} \nn\\
\times\epsilon _1^i\epsilon _2^j\epsilon _3^k e^{-\left( \tau
_1\tau _2\tau _3\right) /\overline{\tau }\left( \sum k_i^2/\tau
_i\right) } \left[ \tau _3\overline{\tau }\left(
k_{1i}k_{2j}k_{1k}-k_{1i}k_{2j}k_{2k}\right) +cyclic\right] \eea

On the other hand, the three-point function from the $AdS$
supergravity, found in (\ref{24}), has the form:
$$
\left\langle J_a^i\left( k_1\right) J_b^j\left( k_2\right)
J_c^k\left( k_3\right) \right\rangle _{f_{abc}}=\frac
i2f_{abc}\delta ^d\left( k_1+k_2+k_3\right) \int\limits_0^\infty
d\tau _1d\tau _2d\tau _3\frac{\pi ^{d/2}}{\overline{\tau
}^{d/2+1}}e^{-\left( \tau _1\tau _2\tau _3\right) / \overline{\tau
}\left( \sum k_i^2/\tau _i\right) }
$$
\begin{equation}
\label{120}\times \left[ \tau _3\overline{\tau }\left(
k_{1i}k_{2j}k_{1k}-k_{1i}k_{2j}k_{2k}\right) +\left( d-2\right)
\tau _1\tau _2\delta _{ij}\left( \tau _3\left(
k_{2k}-k_{1k}\right) +\left( \tau _1-\tau _2\right) k_{3k}\right)
+cyclic\right] .
\end{equation}
Comparing equation (\ref{119}) with the contribution of the first
term in the square bracket of (\ref{120}), we find that they
coincide.

Second term in square bracket (\ref{120}), can be written by means
of $\delta$ -function as well:
$$
\epsilon _1^a\epsilon _2^b\epsilon _3^c\left[ \tau _1\tau _2\delta
_{ab}\left( \tau _3\left( k_{2c}-k_{1c}\right) +\left( \tau
_1-\tau _2\right) k_{3c}\right) +cyclic\right] =
$$
$$
-\overline{\tau }^3\int\limits_0^1d\alpha _k\delta \left( \alpha
_k\right) \sum\limits_{i<j}\alpha _i\alpha _j\epsilon _i\cdot
\epsilon _j\left(
k_i\cdot \epsilon _k\stackrel{.}{G}_B^{ki}+k_j\cdot \epsilon _k\stackrel{.}{G%
}_B^{kj}\right).
$$
As we notice, the $\delta $-function term in
$\stackrel{..}{G}_B^{ij}$ comes from the contribution of tadpole
diagram of scalar loop. It is cancelled with terms of spinor loop
when we sum scalar and spinor contributions. However, the tadpole
diagram of scalar loop was not considered in \cite{21} and so,
here we have got an additional term proportional to
$\delta$-function. So, the second term in (\ref{120}) would be
cancelled with the scalar tadpole contribution, if calculations in
\cite{21} took it into account. This completes the equivalence of
AdS supergravity and gauge theory worldline approaches to the
three point function.

%%%%%%%%%%%%%%%%%%%%%%%%%%%%%%%%%%%%%%%%%%%%%%%%%%%%%%%%%%%%%%%%%%%%%%%%%%%
\Section{Conclusion}

We observed that for the R-current vector fields in the free limit
of the SYM theory, using the worldline formalism, it is possible
to convert the two and three-point functions into the amplitudes
in the supergravity theory in the $AdS$ side. By a convenient
reparametrization, these expressions are written in terms of bulk
to boundary propagators of the corresponding vector field in the
$AdS$. An analogy with  the electrical networks is useful here. In
a Feynman diagram one can consider the moduli parameters
$\tau_i$'s as resistances and the momenta as the electrical
currents in each internal/external line. Now the reparametrization
given in section 4 as $\tau \goes \a$ is equivalent to the
renowned transformation from \emph{delta} (loop) diagram  to
\emph{star} (tree) diagram in the electrical networks. In terms of
string theory worldsheet, it realizes the duality transformation
from the one loop open string to the sphere of closed string. This
is important in proposing a geometrical (in the worldsheet sense)
understanding of the open/closed dualities in string theory.

Finally we have shown that the two and three-point functions
derived from the SYM theory are equivalent to those which come
from the bulk by $AdS/CFT$ correspondence. Although the
three-point function has two independent parts which we denoted by
$f_{abc}$ and $d_{abc}$ in section 1, we have shown the
correspondence for the $f_{abc}$ part only. The $d_{abc}$ part in
the SYM side, comes from the anomaly triangle diagram involving
$\gamma^5$. The worldline treatment of $\gamma^5$ interactions is
well-known and it can be carried on as well. We believe that the
result will match with the $AdS$ calculations, so the open/closed
string duality can be implemented completely.
%%%%%%%%%%%%%%%%%%%%%%%%%%%%%%%%%%%%%%%%%%%%%%%%%%%%%%%%%%%%%%%%
\begin{figure}\label{fig4}
 \begin{center}
  \includegraphics[scale=.6]{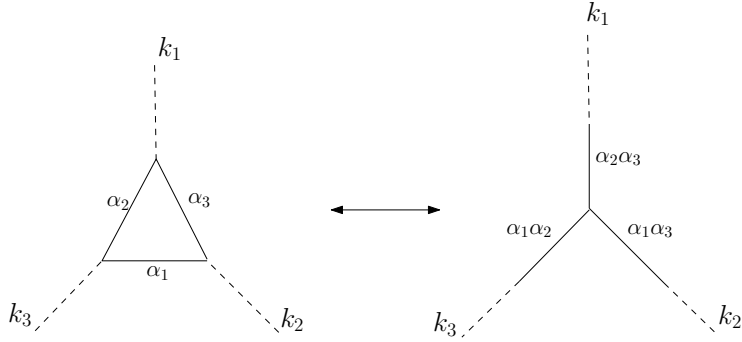}
 \end{center}
\caption{The analogy with the {\it Delta/Star} equivalences in
electrical networks appears in the open/closed duality (compare
with Fig. 1). The momenta are analogous to currents and schwinger
parameters are analogous to resistances.}
\end{figure}

%%%%%%%%%%%%%%%%%%%%%%%%%%%%%%%%%%%%%%%%%%%%%%%%%%%%%%%%%%%%

%%%%%%%%%%%%%%%%%%%%%%%%%%%%%%%%%%%%%%%%%%%%%%%%%%%%%%%%%%%%%%%%%%%%%%%%%%%
%\newpage
\begin{center}
{\large {\bf Acknowledgements}}
\end{center}

%\newline

Sh. M. thanks IPM for financing his visit to this institute. He
acknowledges Prof. F. Ardalan for the invitation and hospitality
during this visit. Authors specially thank R. Gopakumar for
valuable discussions on his work. Also thanks to M. Alishahiha, S.
Azakov, R. Russo, S. Wadia for useful discussions and comments and
A. E. Mosaffa for reading the manuscript. Many thanks to the
members of IPM string theory group for discussions in ISS2005 and
hospitality. S. P. would like to thank the hospitality of the
theoretical physics division at CERN, Geneva, where part of this
work was done. The work of S. P. was supported by Iranian TWAS
chapter Based at ISMO.

%%%%%%%%%%%%%%%%%%%%%%%%%%%%%%%%%%%%%%%%%%%%%%%%%%%%%%%%%%%%%%%%%%%%%%%%%%%%%%%%

\vspace{1cm}
\appendix{{\large{\bf Appendix: The worldline Formalism}}
\vspace{1cm}
\renewcommand{\theequation}{\arabic{equation}}

Here we briefly introduce the worldline approach to the effective
action\footnote{We consider only a scalar loop with external
vector fields. For spinor loop and more details consult with
\cite{15}.}. Let us start from a massive scalar field minimally
coupled to a vector boson: \bea S \sim \int d^d x
\phi^{\dagger}\left[ (\p +i g A)^2 -m^2 \right]\phi \eea The
one-loop effective action then read as, \bea \Gamma[A] &=& -\haf
{\rm Tr} \log \left[ \frac{-(\p
+i g A)^2 +m^2}{-\Box + m^2} \right] \nn\\
&=& \int_0^\infty \frac{d \tau}{\tau} \e^{-m^2 \tau}
\int_{x(\tau)=x(0)} {\cal D}x(\tau)\e^{-\int_0^\tau d\tau \left(
\frac{1}{4} \dot{x}^2 +i g \dot{x}\cdot A(x(\tau))\right)},  \eea
where we use the formula \bea -{\rm Tr} \log \left(
\frac{A}{B}\right) = \int_0^\infty \frac{d\tau}{\tau} {\rm
Tr}\left( \e^{-A\tau} - \e^{-B\tau} \right). \eea

Since we are motivated form the string theory, we can compare the
above effective action with the  bosonic string path integral:
\bea \Gamma[A] &\sim
&\sum_{{\rm top}} \int {\cal D} h {\cal D} x \e^{-S_0 - S_I} \nn\\
S_0 &=& -\frac{1}{4\pi \alpha'} \int_M d\sigma d\tau \sqrt{h}
h^{\alpha\beta} \eta_{\mu\nu} \p_\alpha x^\mu \p_\beta x^\nu
\nn\\
S_I &=& \int_{\p M} d\tau \; i\;g \dot{x}^\mu A_\mu(x(\tau)) .
\eea In the infinite tension limit, $\alpha' \goes 0$, the string
length goes to zero and the string worldsheet approaches to a
worldline. Then if we consider only topology of a single closed
loop as in Fig. 2 with no internal vector field correction it is
comparable with the above effective action.

Now for the effective action with $N$ external vector field we
consider the background as a sum of plane waves with definite
polarizations, \bea A_\mu(x)=\sum_{i=1}^N \epsilon_{i\mu} \e^{i
k_i \cdot x} \eea by expanding and keeping terms linear in each
$\epsilon$ we arrive at, \bea \label{jjjj} \Gamma \sim \langle
\dot{x}_1^{\mu_1} \e^{i k_1\cdot x_1} \cdots \dot{x}_N^{\mu_N}
\e^{i k_N\cdot x_N} \rangle . \eea  This is exactly the amplitude
of N vector vertices in string theory, inserted on a worldline
circle instead of annulus (see Fig. 2).

To derive the expression of the effective action we use the one
dimensional Green function on the loop as the inverse of kinetic
part of the action,
$$ 2 \langle \tau_1 | \left( \frac{d}{d\tau} \right)^{-2} |\tau_2
\rangle = G_B(\tau_1,\tau_2) $$ It can be derived as,
$$G_B(\tau_1,\tau_2)= |\tau_1-\tau_2|
-\frac{(\tau_1-\tau_2)^2}{\tau}\;.$$

The vertex operators in (\ref{jjjj}) can be exponentiate as,
$$  \epsilon_i \cdot \dot{x}_i \e^{i k_i\cdot x_i} = \e^{
\epsilon_i \cdot \dot{x}_i +i k_1\cdot x_1} \mid_{lin{\epsilon_i}}
$$
Then by  completing the square in the path integral and using the
Green function, the effective action with $N$ external vector
legs, in $d=4-\epsilon$ dimension, read as, \bea
 \Gamma _N\left( k_1,...,k_N\right)
&=&\frac{\left( ig\right) ^N}{\left( 4\pi
\right) ^2}{\rm Tr}\left( T^{a_N}\cdots T^{a_1}\right) \int\limits_0^\infty \frac{d\tau}{%
\tau^{3-\epsilon/2}} \e^{-m^2\tau}\prod_{i=1}^N \int d\tau_i  \nn\\
&&\times \e^{ \sum\limits_{i<j=1}^N\left( k_i\cdot
k_jG_B^{ji}-i\left( k_i\cdot \epsilon _j-k_j\cdot \epsilon
_i\right)
\stackrel{.}{G}_B^{ji}+\epsilon _i\cdot \epsilon _j\stackrel{..}{G}%
_B^{ji}\right)}\mid_{linear\ in\ each\ \epsilon } \;, \eea where
we have included the color matrices as well. This result, using
$\tau_i=u_i\tau $ and $m=0$, is equivalent to equation (\ref{28})
which we have used in sections 1 and 2. }
%%%%%%%%%%%%%%%%%%%%%%%%%%%%%%%%%%%%%%%%%%%%%%%%%%%%%%%%%%%%%%%%%%%%%%%%%%%%%%%%

\end{document}